\documentclass[pra,twocolumn,showpacs,preprintnumbers,amsmath,amssymb]{revtex4}
\makeatletter
\usepackage[dvips]{graphicx}	

\usepackage{amsmath}
\usepackage{amssymb}
\usepackage{amsbsy}
\usepackage{color}
\setcitestyle{square}
\usepackage{epstopdf}
\DeclareGraphicsExtensions{.pdf,.eps,.png,.jpg,.mps}

\begin{document}

\title{
Topologically protected gap states and resonances in gated trilayer graphene}
\author{ W. Jask\'olski} \email{wj@fizyka.umk.pl}
\author{G. Sarbicki}
\affiliation{Institute of Physics, Faculty of Physics, Astronomy and Informatics, Nicolaus Copernicus University in Torun, Grudziadzka 5, 87-100 Toru\'n, Poland}

\date{\today}

%%%%%%%%%%%%%%%%%%%%%%%%%%%%%%%%%%%%%%%%%%%%%%%%%%%%%%%%%%%%%%%%%%%%%%%%%%%%

\begin{abstract} 
Gated  trilayer graphene exhibits energy gap  in  its most stable ABC stacking. Here we  show  that when  the  stacking  order  changes  from ABC  to  CBA,  three  gapless states  appear  in  each  valley.  The  states  are  topologically  protected  and  their  number  is  related  to  the  change  of  the  valley Chern  number  across  the  stacking boundary.  The stacking change  is achieved by corrugation or delamination  in  the top  and  bottom  layers, which  simultaneously  yields  two  AB/BA  stacking  domain walls in the pairs of adjacent layers (in bilayers). This in turn causes that for some gate  voltages  two pairs of  
topological resonances appear additionally in the conduction and valence band continua.  
\end{abstract}

\pacs{73.63.-b, 72.80.Vp}

\maketitle

%%%%%%%%%%%%%%%%%%%%%%%%%%%%%%%%%%%%%%%%%%%%%%%%%%%%%%%%%%%%%%%%%%%%%%%%%%%%%%% INTRODUCTION %%%
%%%%%%%%%%%%%%%%%%%%%%%%%%%%%%%%%%%%%%%%%%%%%%%%%%%%%%%%%%%%%%%%%%%%%%%%%%%

\section{\label{sec:intro} Introduction}

%{\it{Introduction.}} 
Bilayer graphene (BLG) with AB stacking order of sublattices opens tunable energy gap under applied gate voltage 
\cite{Ohta_2006,Castro_2007,Oostinga_2008,Zhang_2009}.
Similarly, gated trilayer graphene (TLG) also shows energy gap in its most stable ABC stacking \cite{Zhang_PRB_2010,Liu_np_2011}. 
Both systems focus attention as possible candidates for graphene-based electronics 
\cite{Schwierz_2010,Choi_2010,Santos_2012,Zhang_transistor_2018}, also
due to superconductivity reported recently in BLG and TLG  with twisted layers \cite{Jarillo2018,Chen_Nature_2019,Chittari_PRL_2019}.

Gated bilayer graphene presents an interesting property: when the stacking order changes from AB to BA, topological valley-protected  states appear in the energy gap, connecting the conduction and valence  band continua \cite{Vaezi_2013, Zhang_2013,Alden_2013,San_Jose_2014,Pelc_2015,Yin_NC_2016}. These states are localized at the stacking domain wall (SDW) and can provide one-dimensional currents along the SDW \cite{Ju_2015,Li_2016}. The stacking change may occur in the BLG when one layer is stretched, corrugated or delaminated \cite{Lin_2013,Ju_2015,Pelc_2015,Peeters_2018}, or even when it contains grain boundaries \cite{Jaskolski_2016}.
Valley-protected states occur also in the gap of BLG and TLG when the gate polarity changes sign \cite{Martin_2008,Zhang_2013,Jung_PRB_2011}.

In this paper we investigate the electronic structure of gated trilayer graphene with stacking domain walls. We demonstrate that when the stacking order changes from ABC to CBA, three topologically protected states appear in each valley. This result is obtained by calculating the local density of states (LDOS) at stacking boundaries. Then it is confirmed by the analysis of the valley Chern number. We calculate its value and show that it 
changes by three when the stacking of trilayer graphene changes from ABC to CBA. We also demonstrate that for some values of the gate voltage, topologically protected resonances appear in the conduction and valence band continua, with properties characteristic for gapless states of gated bilayer with AB/BA stacking domain wall.

Two upper or two lower layers of TLG can be seen as graphene bilayers. If AB/BA stacking change is created in one of these bilayers by, for example, stretching or corrugating the upper layer only, the stacking of TLG changes from ABC to ABA. However, trilayer graphene in the ABA stacking is not gapped even under gate voltage \cite{Koshino_2009,Zou_nl_2013,Zhang_2013}.
To achieve ABC/CBA stacking change in TLG, the AB/BA-type stacking domain walls have to be created in both bilayers. This can be achieved by simultaneous stretch or corrugation (delamination) in both, the lower and the upper layers of TLG.

It is of importance to recognize multilayer graphene systems gapped under applied electric field with topological gapless states. Such states can provide one-dimensional currents along defined directions and thus can be useful in graphene-based electronics. 
%

%% FIGURE %%%
\begin{figure}[ht]
\centering
\includegraphics[width=\columnwidth]{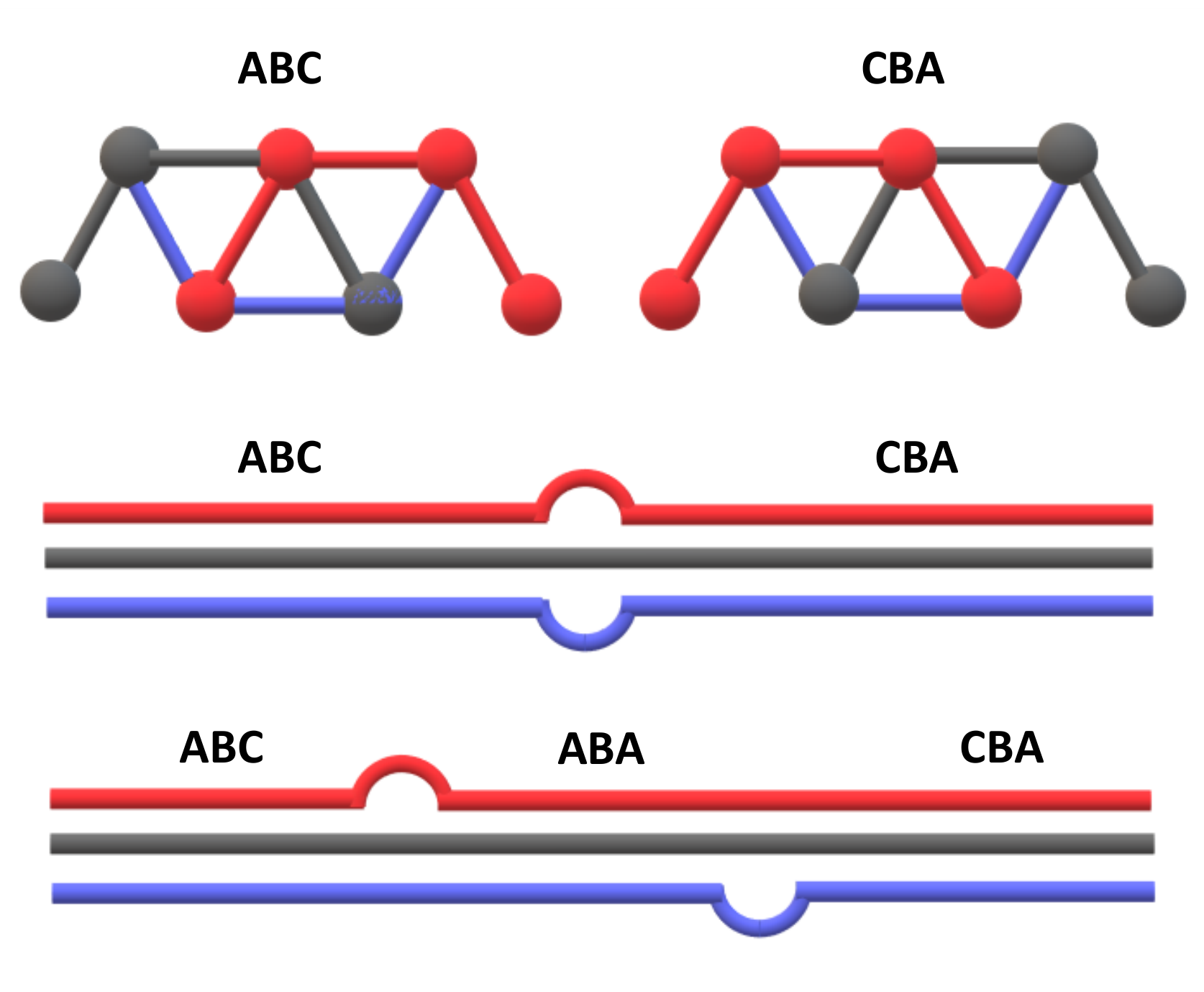}
\caption{\label{fig:first}
Top: schematic representation of two different stackings of trilayer graphene, ABC and CBA. 
Middle and bottom: stacking change created by corrugation or delamination  in the upper and lower layers along the zigzag direction - horizontal view along the armchair direction. Corrugation  in the upper layer creates AB/BA stacking domain wall in the upper bilayer (UBSDW). Corrugation  in the lower layer creates AB/BA stacking domain wall in the lower bilayer (LBSDW). When the LBSDW and UBSDW are spatially separated, the trilayer ABA stacking appears in between.
}
 \end{figure}

%% FIGURE %%%
\begin{figure}[ht]
\centering
\includegraphics[width=\columnwidth]{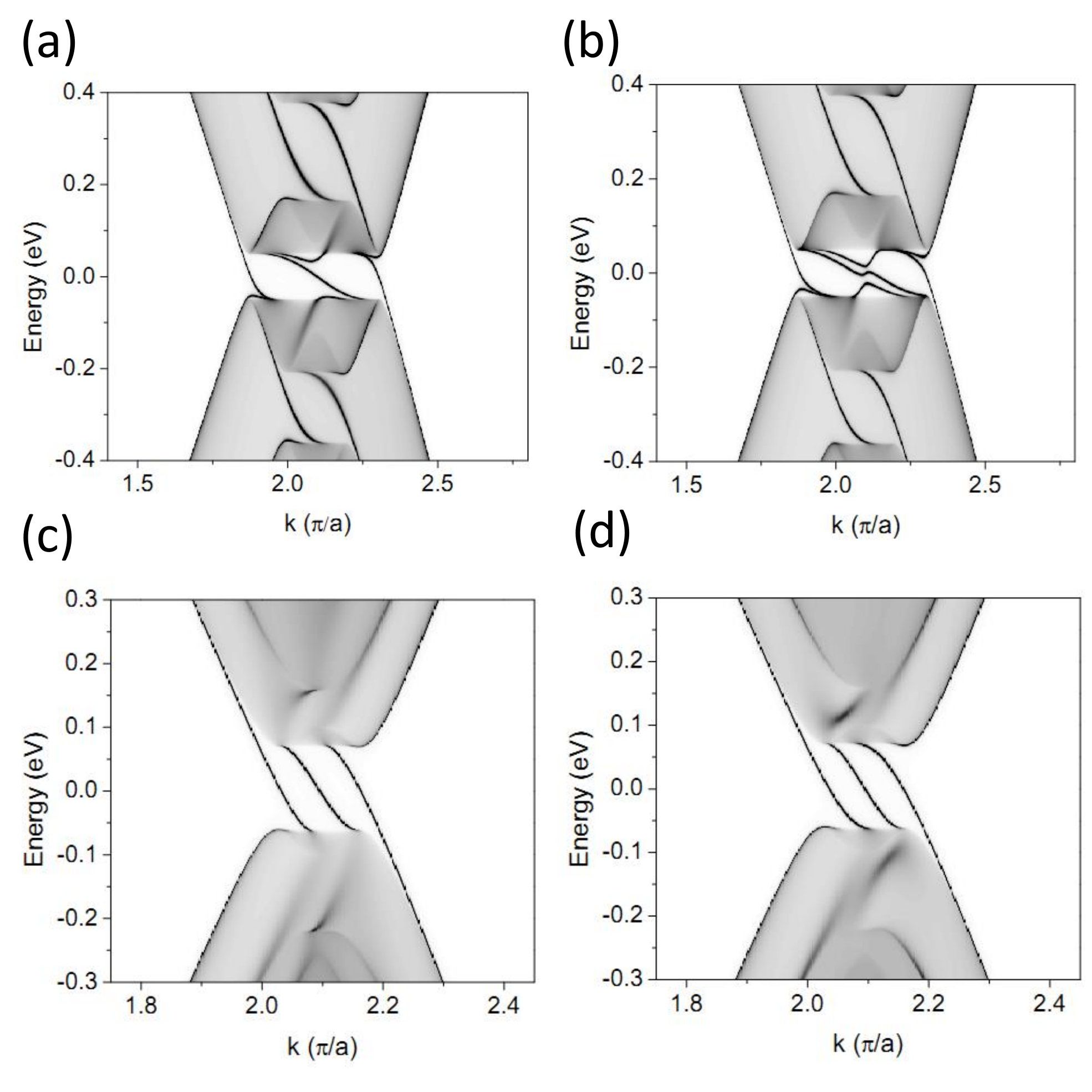}
\caption{\label{fig:second}
Local density of states calculated along the zigzag direction in the region shown in Fig. \ref{fig:first}, for two different gate voltages applied between the upper and the lower layers. (a) and (b) - $V=\pm 0.5$ eV, (c) and (d) - $V=\pm 0.1$ eV. (a) and (c) - LBSDW and UBSDW overlap, (b) and (d) - LBSDW and UBSDW separated.
}
 \end{figure}

\section{Model and methods}

%{\it{Model and Methods}}. 
We investigate trilayer graphene in which  the upper and the lower layers are corrugated, as schematically shown in the middle and bottom panels of Fig. \ref{fig:first}. In the upper panel the 12-atom unit cells of ABC and CBA trilayers are shown. The corrugations come down to delamination of one layer versus the neighboring one \cite{Pelc_2015}, what leads to the AB/BA stacking change in a given bilayer. We call stacking domain wall in the upper bilayer as UBSDW, and stacking domain wall created in the lower bilayer as LBSDW. The presence of UBSDW and LBSDW leads automatically  to the stacking change between the upper and lower layer, although they are not directly connected. We consider the smallest corrugations possible, i.e., delaminating only one unit cell in the lower layer and half a unit cell in the upper layer. It was shown in Ref.\cite{Pelc_2015} that the appearance of topological gapless states is determined by the presence of SDW, not by the size of the corrugation. The system is infinite in both, the armchair and the zigzag directions, the corrugation and thus SDW are along the zigzag direction.

%% FIGURE %%%
\begin{figure}[ht]
\centering
\includegraphics[width=\columnwidth]{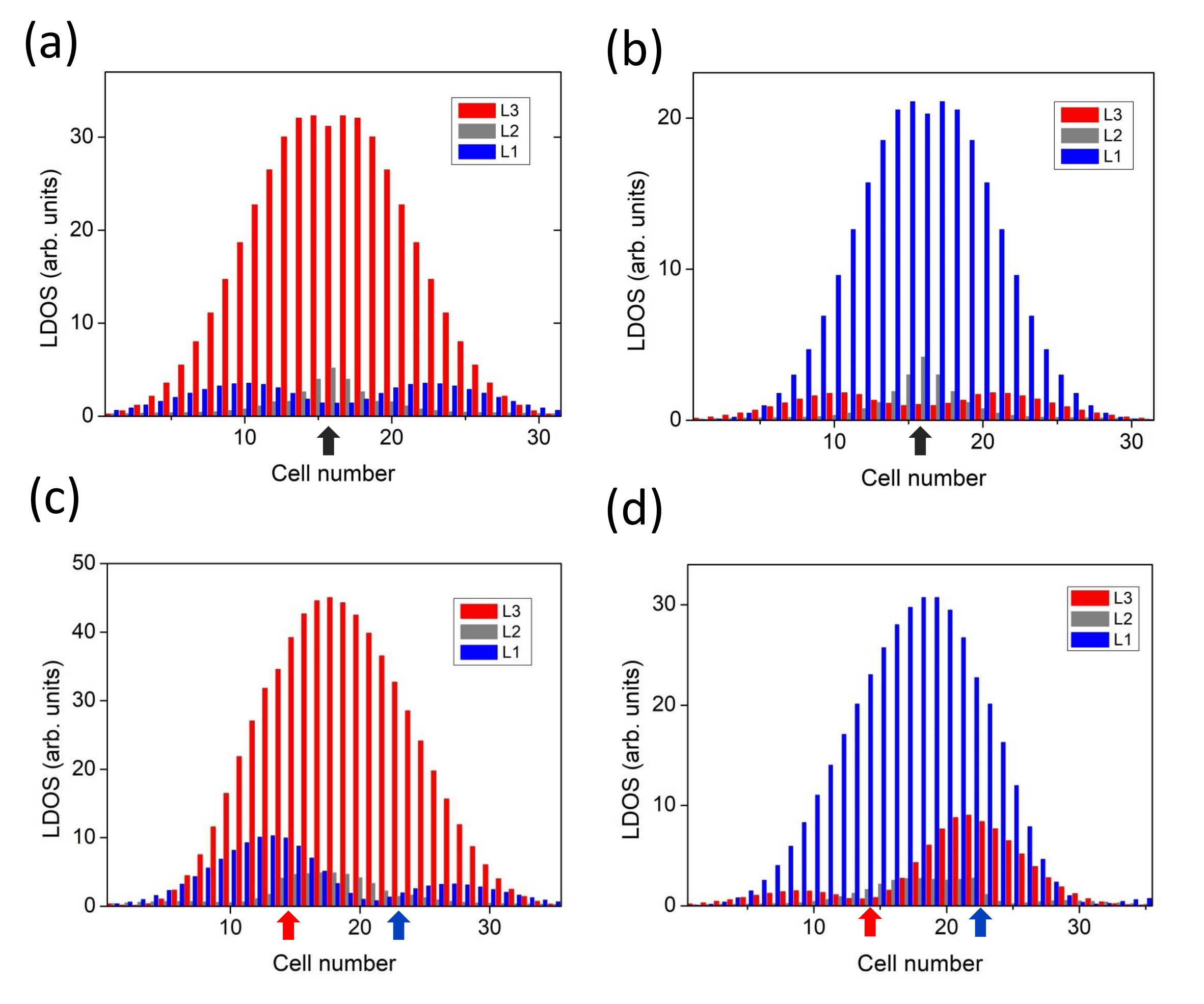}
\caption{\label{fig:third} 
Layer-resolved density distribution of the gapless states (at $E=0$) shown in Fig. \ref{fig:second}  (a) and (b), for gate voltage $V= \pm 0.5$ eV. (a) and (b) – left and right gap states for the case when UBSDW and LBSDW spatially overlap (Fig. \ref{fig:second}  (a)); (c) and (d) -  left and right gap states for the case when UBSDW and LBSDW are spatially separated (Fig. \ref{fig:second}  (b)).
Layer densities: lower (L1) – blue, middle (L2) – grey, upper (L3) – red. Vertical bars represent  densities calculated at 4-atom unit cells of each layer, shown in Fig. \ref{fig:first}. Black arrow marks the position of LBSDW and UBSDW when they overlap. Red and blue arrows mark the position of separated UBSDW and LBSDW, respectively.
}
\end{figure}

Two different cases are investigated, (a) when UBSDW and LBSDW spatially overlap, and (b) when they are separated by seven unit cells, as shown in Fig. \ref{fig:first}. 
In the latter case, ABA trilayer is formed between both SDWs. In both cases we study how the structure of gapless states depends on the value of the gate voltage. 

We work within the $\pi$-electron tight binding approximation. Intra-layer and inter-layer hopping parameters $\gamma_0=2.7$ eV and $\gamma_1 = 0.27$ eV are used, respectively \cite{Castro_2007,Ohta_2006}. Voltages $+V$ and $-V$ are applied to the bottom and top layers, respectively.  
Two values of $V$ are taken into account, $V < \gamma_1$ and $V > \gamma_1$. This choice is dictated by the observation that topological gapless states in BLG with stacking domain walls localize in different layers depending on the value of $V$ vs $\gamma_1$ \cite{Jaskolski_2018}.
Local density of states along the zigzag direction is calculated for the region containing both UBSDW and LBSDW and several unit cells to the left and to the right from the corrugations (as shown in  Fig. \ref{fig:first}). To calculate LDOS we use the surface Green function matching technique \cite{Datta,Nardelli_1999}.

\section{Results}
\subsection{Numerical calculations}
%{\it{Results: numerical calculations}}.
In Fig. \ref{fig:second} we present local density of states calculated for TLG with two minimal corrugations in the lower and upper layers, as shown in the middle and bottom panels of Fig. \ref{fig:first}. The corrugations cause that the stacking of TLG changes from ABC (left side) to CBA (right side).
Since TLG is periodic in the zigzag direction (along the corrugations) the LDOS is $k$-dependent, where $k$ is the wavevector corresponding to this periodicity. In the upper panels of the figure, gate voltages   applied to external layers are  $V=\pm 0.5$ eV, while in the lower panels $V=\pm 0.1$ eV. The left panels correspond to the case of spatially overlapping LBSDW and UBSDW, right panels concern separated LBSDW and UBSDW.
In all cases there are three states connecting the valence and conduction bands across the energy gap. The figure shows LDOS calculated for $k$ close to the cone at $K$; for $K^{\prime}$ the figure is mirror-reflected, so the gap states are valley-protected.

%% FIGURE %%%
\begin{figure}[ht]
\centering
\includegraphics[width=\columnwidth]{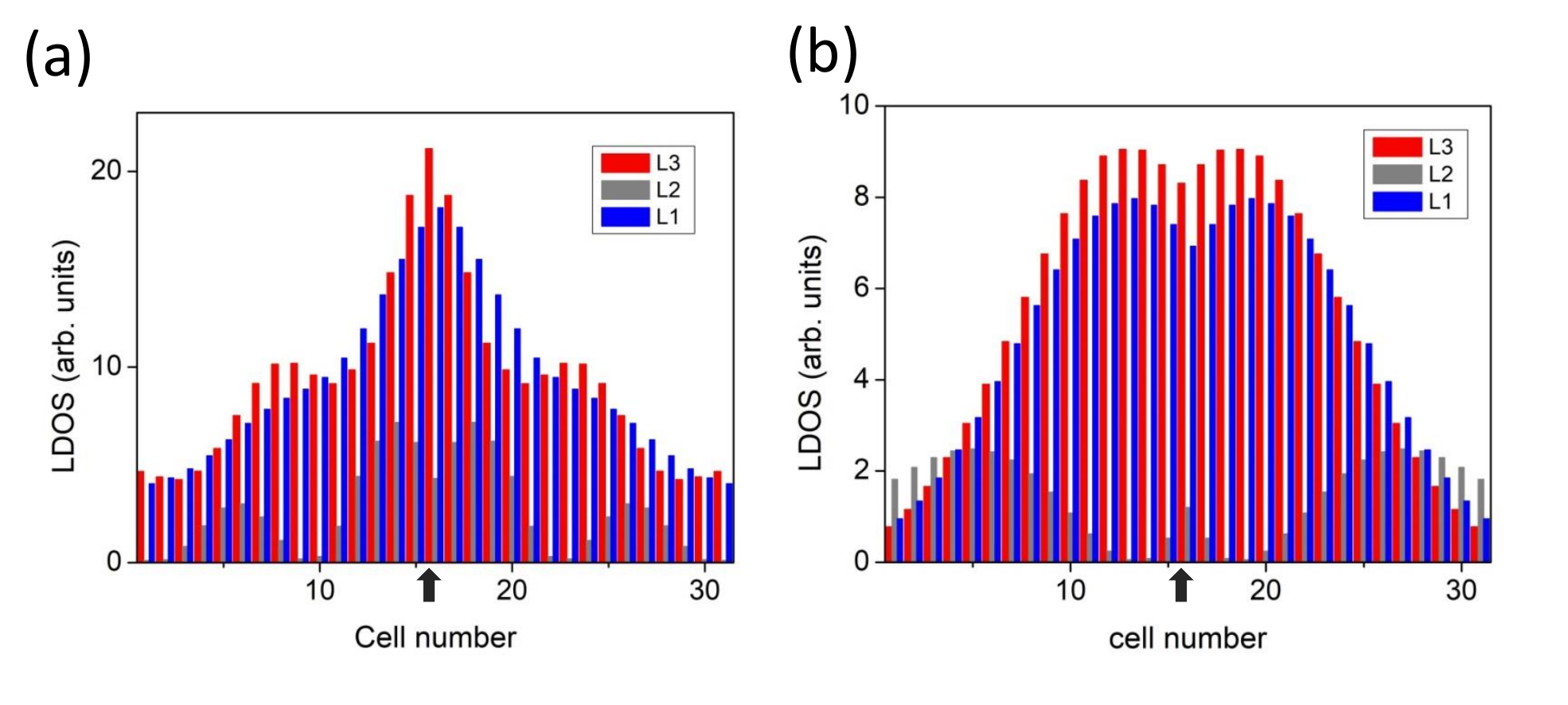}
\caption{\label{fig:fourth} 
Layer-resolved density distribution of the middle gapless states shown in Fig. 2 a and c,  i.e., for gate voltages $V= \pm 0.5$ eV and  and $V= \pm 0.1$ eV, for overlapping LBSDW and UBSDW. Colors and arrows as in Fig. \ref{fig:third}.
}
\end{figure}

Now, we analyze the spatial localization of the gapless states. We do it first for the case of $V=\pm 0.5$ eV.
In Fig. \ref{fig:third} (a) and (b) we present the layer-resolved LDOS calculated for $k$-values for which $E=0$ line (the Fermi energy) crosses the left and right gapless states seen in Fig. \ref{fig:second} (a), i.e., for the case of overlapping LBSDW and UBSDW. The envelope of the vertical bars reflects the density of the corresponding wave function. For this gate voltage the states are almost entirely localized only in the lower or in the upper layer, just in place where the UBSDW and LBSDW occur. 
When the corrugations are separated (panels (c) and (d)) the localization of these states moves to the area between UBSDW and LBSDW. 
The LDOS for the middle gap state, presented in Fig. \ref{fig:fourth} (a), is almost equally distributed between the lower and the upper layers, with small contribution from the middle layer \cite{note2}. We can conclude that for this gate voltage the gapless states are due to stacking change between these two layers, which we call {\it indirect stacking}, since their nodes are not connected by $\gamma_1$.

%% FIGURE %%%
\begin{figure}[ht]
\centering
\includegraphics[width=\columnwidth]{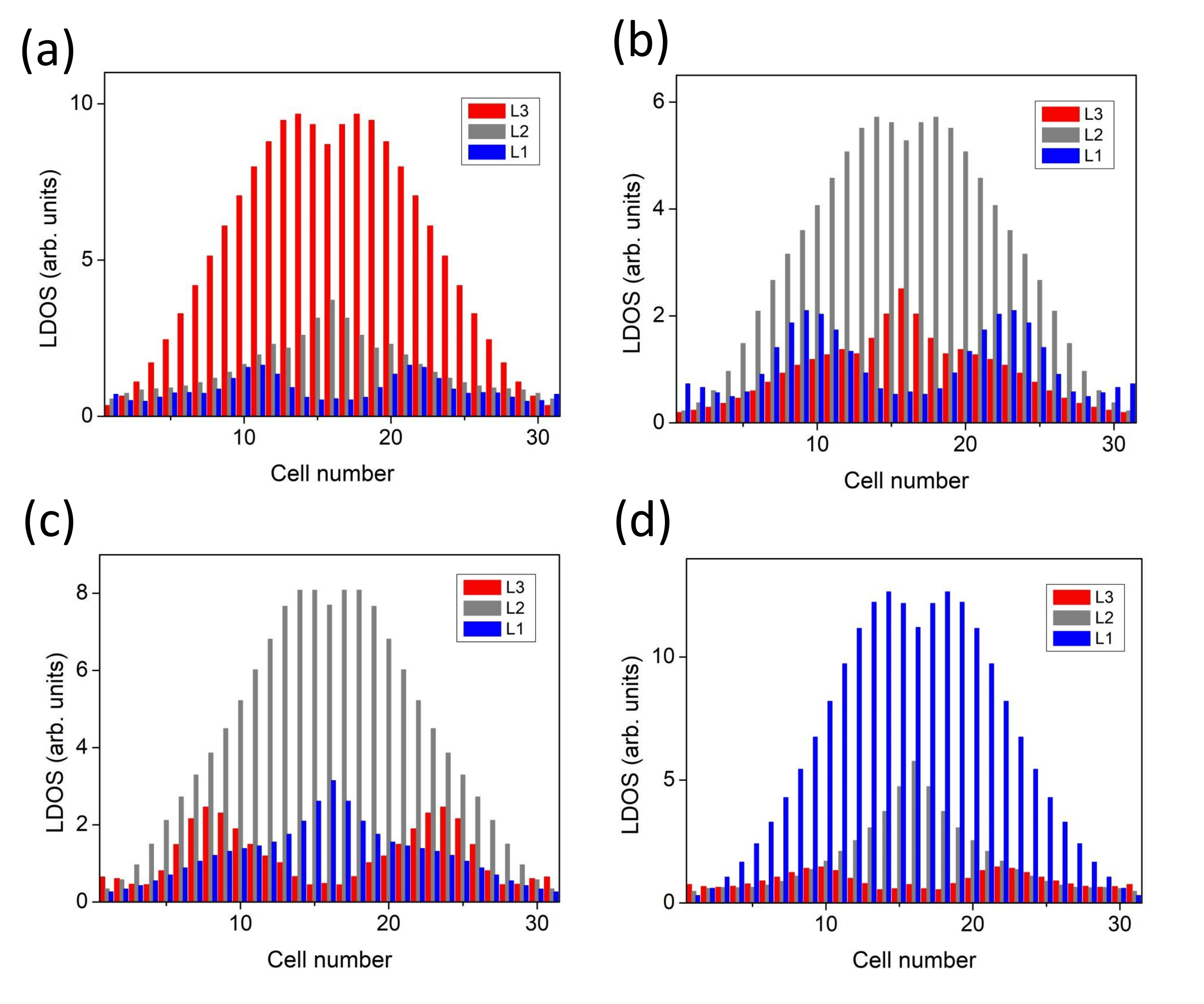}
\caption{\label{fig:fifth} 
Layer-resolved density distribution of the gapless resonances shown in Fig. \ref{fig:second}  (a), i.e., for the gate voltage $V=\pm 0.5$ eV. (a) and (b) - valence band resonances calculated for energy  $E=-0.29$ eV. (c) and (d) - conduction band resonances calculated for energy $E=0.28$ eV.
}
\end{figure}

Closer inspection of Fig. \ref{fig:second} (a) and (b) reveals also two resonance states below and above the Fermi level, i.e., in the valence and conduction band continua. The energy structure of these resonances reminds topological gapless states of bilayer graphene \cite{Zhang_2013,Pelc_2015}. Indeed, the corresponding layer-resolved LDOS, shown in Fig. \ref{fig:fifth}, confirm this supposition. The left resonance state in the valence band is localized in the upper layer, while the right state is localized in the middle layer, just like the gapless states of the gated upper bilayer with UBSDW \cite{Jaskolski_2018}. In the case of resonances situated in the conduction band, the left one is localized in the middle layer and the right one in the bottom layer, again, like the gapless states of the lower bilayer with LBSDW and inverted gate polarization. 
%
% WJ – resonances explanation
%
The appearance of resonances can be explained as follows.  
The gate $V > \gamma_1$ moves the electron charge 
from the bottom to the middle layer, and from the middle layer 
to the top one. We thus get the bottom gated bilayer with the 
lower layer discharged - yielding two topological states 
in the unoccupied conduction band, 
and the upper gated bilayer with the top layer additionally loaded - 
yielding a couple of topological states in the valence band continuum. 
When $V < \gamma_1$ the electron charge is not moved between the layers 
and the resonances do not appear (Fig. \ref{fig:second} c,d).

%% FIGURE %%%
\begin{figure}[ht]
\centering
\includegraphics[width=\columnwidth]{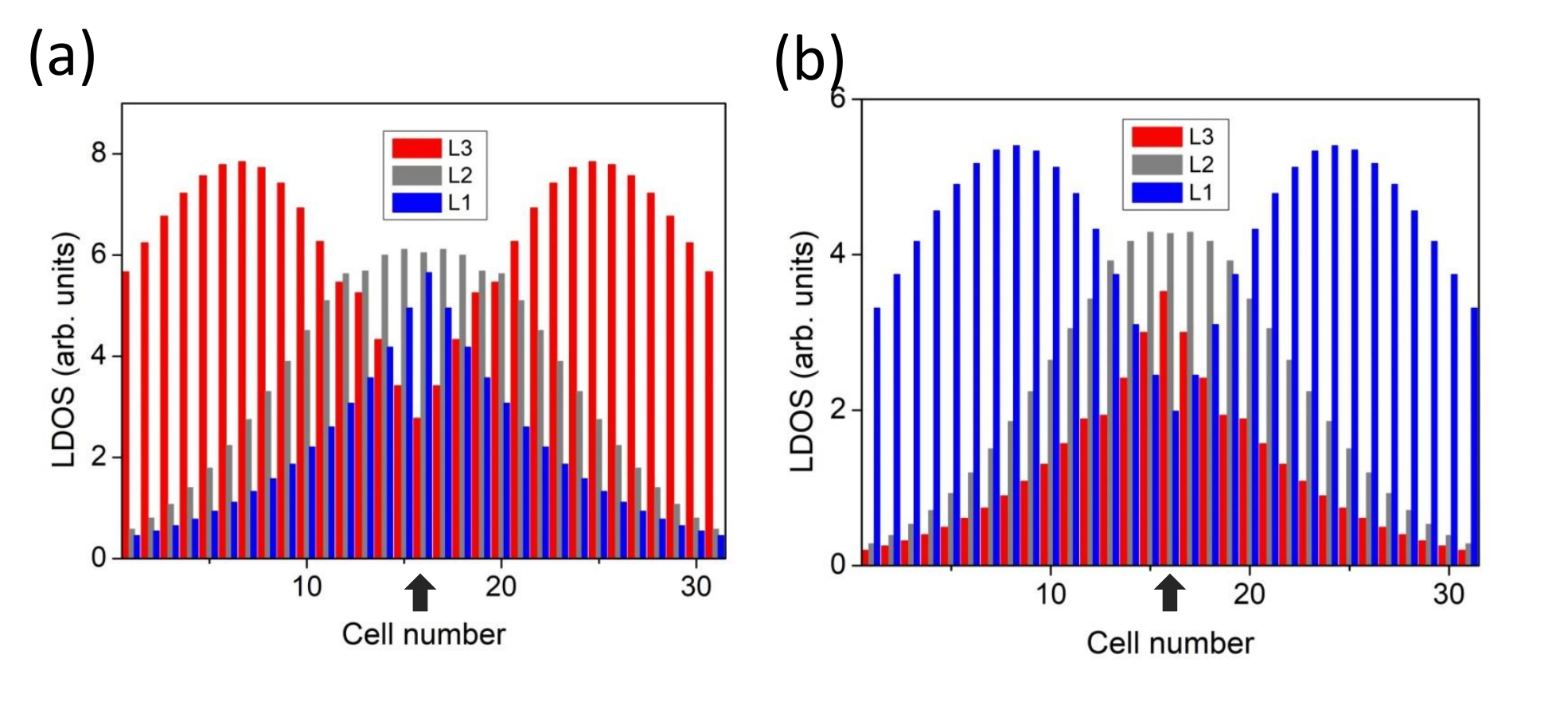}
\caption{\label{fig:six} 
Layer-resolved density distribution of the gapless states (at $E$=0) shown in Fig. \ref{fig:second}  (c), i.e., 
for gate voltage $V= \pm 0.1$ eV and spatially overlapping LBSDW and UBSDW. (a)  left state, (b) right state. 
}
\end{figure}

The triplet of gapless states occurs for any gate voltage. However, for $V < \gamma_1$, their localization is different. The layer-resolved LDOS for the left and right gap states seen in Fig. \ref{fig:second} (c) are presented in Fig. \ref{fig:six} (a) and (b), respectively; the layer-resolved LDOS of the middle gap state is visualized in Fig. \ref{fig:fourth} (b).  For $V=\pm 0.1$ eV the left and right gapless states are more diffused and more evenly distributed between all three layers. 
We have checked that when UBSDW and LBSDW are separated, the layer mixing and diffusion are even stronger.
Now, the gap states do not have the character of bilayer topological states, but reflect the global ABC/CBA stacking change in TLG.

%{\it{Results: analysis of the Chern number}}.

\subsection{Analysis of the valley Chern number}

We begin with the Hamiltonian for trilayer graphene. Unlike the commonly applied reduced Hamiltonians for $N$-layer graphene \cite{Min_2008,Jung_PRB_2011,Chittari_PRL_2019}, we use the full TLG Hamiltonian in its most general form. 
This enables direct analysis of its symmetry when the stacking changes from ABC to CBA. 
The final formulas show how many terms in the eigenvalue expansion are sufficient for the calculation of of the valley Chern number.
For momenta close to the Dirac point $K$, 
the Hamiltonian $H_{ABC}$ of trilayer graphene is a $3 \times 3$ matrix of $2 \times 2$ blocks:
\begin{equation} \label{Hamilt}
 H_{ABC}= \left(
\begin{array}{c|c|c}
  {\rm {\bf \Pi}} + V {\rm {\bf I_2}} & {\rm {\bf T}} & {\rm {\bf 0}} \\ \hline
  {\rm {\bf T}}^T & {\rm {\bf \Pi}} & {\rm {\bf T}} \\ 
\hline
  {\rm {\bf 0}} & {\rm {\bf T}}^T & {\rm {\bf \Pi}} - V {\rm {\bf I_2}}
\end{array} \right),
\end{equation}
where ${\rm {\bf \Pi}} = \nu \left( p_x {\rm {\bf \sigma_x}} +  p_y {\rm {\bf \sigma_y}} \right)$, ${\rm {\bf T}} = \gamma_1 \frac 12 \left( {\rm {\bf \sigma_x}} + i {\rm {\bf \sigma_y }}\right)$ and ${\rm {\bf I_2}}$ is the $2\times 2$ unit matrix,
$\nu= \frac{\sqrt{3}a}{2 \hbar} \gamma_0 $ is the Fermi velocity, and $a$ is graphene lattice constant \cite{Min_2008}. 
%Its lowest energy, $\lambda_0$,  eigenvector is
%\textcolor{red}
Its (unnormalised) eigenvector $\Psi_0$ related to the lowest energy $\lambda_0$ is
\begin{equation}
  \Psi_0 = 
  \left[ \begin{array}{c}
    -\gamma_1V_-(\lambda_0(P^2-V_+^2)+\gamma_1^2V_+) \\
    \gamma_1P\exp(i\phi)(\lambda_0(P^2-V_+^2)+\gamma_1^2V_+) \\
    -P\exp(-i\phi)(P^2-V_-^2)(P^2-V_+^2) \\
    -(P^2-V_-^2)(\lambda_0(P^2-V_+^2) -\gamma_1^2 V_+) \\
    \gamma_1P^2\exp(-2i\phi)(P^2-V_-^2) \\
    \gamma_1V_+P\exp(-i\phi)(P^2-V_-^2)
  \end{array} \right],
\end{equation}
where we use polar coordinates: $\nu(p_x+ip_y) = P\exp(i\phi)$
%, $\phi=\arctan(p_y / p_x)$
and $V_{\pm} = V \pm \lambda_0$.

The valley Chern number $N_\nu = \gamma / 2\pi$ at $K$ is defined by the Berry phase \cite{Jung_PRB_2011,Chittari_PRL_2019}
\begin{equation} \label{Berry}
  \gamma = \mathrm{Im} \oint_C \frac{\langle \Psi_0 | \nabla \Psi_0 \rangle}{\langle \Psi_0 | \Psi_0 \rangle},
\end{equation}
with the integration performed over a circle  $S_1(K,P)$. 
The calculation of $\mathrm{Im} \langle \Psi_0 | \nabla \Psi_0 \rangle$ yields $( \gamma_1^2(\lambda_0(P^2-V_+^2)+\gamma_1^2V_+)^2 -(P^2-V_-^2)^2(P^2-V_+^2)^2-\gamma_1^2(P^2 - V_-^2)^2(V_+^2 + 2P^2) ) P^2 \mathrm{d}\phi$ and $|\Psi_0|^2 = (\lambda_0(P^2-V_+^2)+\gamma_1^2V_+)^2
    (\gamma_1^2V_-^2+\gamma_1^2P^2+(P^2-V_-^2)^2)
    + P^2 (P^2-V_-^2)^2
    ( (P^2-V_+^2)^2 + \gamma_1^2P^2 +\gamma_1^2V_+^2 )$.
To remove singularities, which appear when $P \to \infty$, 
we use the Rayleigh-Schr\"odinger perturbation theory (with the 
zero-order Hamiltonian 
${\rm {\bf I_3 \otimes \Pi}}$). 
We account up to the second correction (of the order $P^{-1}$) and get that $\lambda_0 = -P - V + \frac{\gamma_1^2}{8P} + o(P^{-2})$.
The leading term in $\mathrm{Im} \langle \Psi_0 | \nabla \Psi_0 \rangle$ 
gives $-96\pi \gamma_1^2V^2P^6$, while the leading term in $|\Psi_0|^2$ is equal $32 \gamma_1^2V^2P^6$. Hence, the Berry phase is equal to $-3\pi$ and the contribution to the Chern number from $K$ is $-3/2$ \cite{Jung_PRB_2011,Chittari_PRL_2019}.

Switching from $K$ to $K'$ causes transposition of the matrix ${\rm {\bf \Pi}}$ in the Eq. (\ref{Hamilt}), which is equivalent to parity change 
and yields the opposite sign of the valley Chern number at $K'$. The switching from ABC to CBA stacking causes transposition of the matrix ${\rm {\bf T}}$.
Composition of these switchings is equivalent to a unitary symmetry 
${\rm {\bf I_3}} \otimes {\rm {\bf \sigma_x}}$ on the Hamiltonian. 
This means that in case of CBA stacking, the contributions from neighbourhoods of $K$ and $K'$ to the Chern number are $3/2$ and $-3/2$, respectively, i.e., they have reversed signs comparing to the ABC stacking. Therefore, the stacking change causes change of the local Chern number contribution by $\pm 3$ and thus the appearance of three gapless states localized at the stacking domain wall.

It is worth noting that $H_{CBA}$ can be unitarily transformed to 
$H_{ABC}(-V)$, i.e., with the sign of $V$ reversed. This means that three gapless states appear also when the gate polarization changes, as was shown in Ref. \cite{Jung_PRB_2011}.

\section{Conclusions}

%{\it{Conclusions}}.
We have investigated gated trilayer graphene in the ABC stacking. We have shown that by simultaneous corrugation or delamination in the upper and lower layers, the stacking may change from ABC to CBA. This change automatically creates AB/BA-type stacking changes in the upper bilayer formed by the upper and middle layers, in the lower bilayer formed by the middle and lower layers, as well as between the upper and the lower layer, which is the indirect stacking change, since these layers are not directly connected by the interlayer hopping. 

We have calculated the LDOS along the stacking domain walls and  demonstrated the presence of three valley-protected topological states in the energy gap. We have analyzed the valley Chern number for TLG and shown that it changes from -3/2 to 3/2 when the stacking changes from ABC to CBA, yielding three gapless states in the energy gap at each valley. 

The gapless states are robust and independent on the voltage $V$ applied to the external layers, but their distribution between the layers depends on $V$. 
When $V$ is greater than the interlayer hopping $\gamma_1$ the gapless states localize almost exclusively in the external layers, what means  that they are due to indirect stacking change between these layers. 
For this voltage two pairs of topological resonance states appear additionally  in the valence and conduction bands, with the characteristics typical for bilayers with SDW.
When $V < \gamma_1$ the resonance states disappear and the gapless states are more evenly distributed between all three layers. We can conclude that one-dimensional currents provided by the gap states can flow in the surface layers only, or in the entire trilayer graphene, depending on the gate voltage applied.

The appearance of topological gapless states and resonances in gated trilayer graphene with stacking change has not been reported so far. We strongly believe that these findings are important both for better understanding of the fundamental features of multilayer graphene, and for potential applications of such systems in graphene-based electronics.

%\bibliography{bib_2018}

\begin{references}
\bibitem{note2} Fig. \ref{fig:second} b reveals a couple of states detaching from the valence and conduction band continua.
They originate from the ABA section of TLG, which is gapless, and appear faster in the gap when $V > \gamma_1$.
\end{references}

%\end{document}

\end{document}